\documentclass[10pt,conference,dvipsnames]{IEEEtran}

\IEEEoverridecommandlockouts
\def \mytype{}
\def \final{}
\def \print{}
\usepackage{textcomp}
\usepackage[utf8]{inputenc}
\usepackage[T1]{fontenc}
\ifx \mytype \empty%
\usepackage{color}
\usepackage[rgb]{xcolor}
\else%
\fi%
\usepackage{listings}
\usepackage{eso-pic} 
\usepackage{graphicx}
\usepackage{mathptmx}
\usepackage{tabularx}
\usepackage{ragged2e}
\usepackage[singlelinecheck=false]{caption}
\usepackage[binary-units=true]{siunitx}
\usepackage{amsmath,amssymb,amsfonts}
\usepackage{nicefrac}
\usepackage{amsmath}
\ifx \mytype \empty%
\usepackage[%
  natbib=true,%
  style=numeric,%
  sorting=none,%
  backend=biber,%
  maxnames=5,%
  mincitenames=1,%
  maxcitenames=2,%
  language=american,%
  firstinits=true,%
  doi=false,%
  isbn=false]{biblatex}
\else%
\usepackage[backend=biber,sorting=none,doi=true,style=ieee]{biblatex} 
\fi%
\usepackage[ruled,linesnumbered]{algorithm2e}
\usepackage{pgf}
\usepackage{pgfplots}
\usepackage{pgfplotstable}
\usepgfplotslibrary{units}
\pgfplotsset{compat=newest} 
\usepackage{tikz}
\usetikzlibrary{shapes,arrows,backgrounds,calc,decorations.pathreplacing,patterns,positioning,fit,shadows.blur,tikzmark}
\ifx \mytype \empty%
\usepackage[caption=false]{subfig}
\usepackage{wrapfig}
\else%
\usepackage[caption=false,font=normalsize,labelfont=sf,textfont=sf]{subfig}
\usepackage{wrapfig}
\pgfplotsset{compat=newest}
\fi%
\usepackage{todonotes}
\ifx \mytype \empty%
\usepackage[inline]{enumitem}
\else%
\fi%
\usepackage{booktabs}
\usepackage{multirow}
\usepackage{acronym}
\usepackage[abspath]{currfile}
\ifx \mytype \empty%
\usepackage[%
  \ifx\print\empty%
    draft,%
    hidelinks,%
    bookmarks=false%
  \else%
    colorlinks%
  \fi]{hyperref}
\else%
  \ifx \mytype poster%
  \else%
  \fi%
\fi%
\usepackage{cleveref}
\Crefname{figure}{Fig.}{Figs.}
\usepackage{lipsum}
\usepackage{tcolorbox}

\makeatletter
\def\beamer@calltheme#1#2#3{%
\def\beamer@themelist{#2}
\@for\beamer@themename:=\beamer@themelist\do
{\usepackage[{#1}]{\beamer@themelocation/#3\beamer@themename}}}

\def\usefolder#1{
\def\beamer@themelocation{#1}
}
\def\beamer@themelocation{}

\xspaceaddexceptions{\,}
 
\newcommand{\ie}{\mbox{i.\,e.,}\xspace}
\newcommand{\eg}{\mbox{e.\,g.,}\xspace}

\input glyphtounicode.tex
\crefname{equation}{eq.}{eqs.}
\Crefname{equation}{Eq.}{Eqs.}

\let\cref\Cref

\graphicspath{{../figures/}}

\usetikzlibrary{external}
\tikzset{external/mode=list and make}

\usepackage{calc}
\usepackage{listings}
\usepackage{xcolor}
\usepackage{scalefnt}

\lstdefinelanguage{HIPAcc}{%
  language=C++,
  morekeywords={size_t, uchar, ushort, uint, uchar4, char4, ushort4, short4,
    uint4, int4, float4, double4},
  keywords=[2]{Image, Accessor, Kernel, IterationSpace, Mask, BoundaryCondition,
    Boundary, CLAMP, MIRROR, UNDEFINED, REPEAT, CONSTANT, output, convolve,
    iterate, reduce, Reduce, SUM, MIN, MAX, PROD, MEDIAN Domain},
  keywordstyle=[2]{\color{Cerulean}\bfseries}
}

\lstdefinestyle{CUDA}{%
  language=C++,
  morekeywords={size_t, uchar, ushort, uint, uchar4, char4, ushort4, short4,
    uint4, int4, float4, double4},
  otherkeywords=[2]{>>>,<<<,[,]},
  alsoletter=[2]{.},
  keywords=[2]{__device__, __host__, __global__, __constant__, __shared__, dim3,
    blockIdx.x, blockIdx.y, blockDim.x, blockDim.y, blockDim.z, threadIdx.x,
    threadIdx.y, threadIdx.z, gridDim.x, gridDim.y, warpSize, __syncthreads,
    __sinf, __cosf, __expf, cudaMalloc, cudaMemset, cudaFree, cudaMemcpy,
    cudaMemcpyHostToDevice, cudaMemcpyDeviceToHost, cudaMemcpyDeviceToDevice,
    tex1Dfetch, tex2D, surf2Dwrite},
  keywordstyle=[2]{\color{blue}\bfseries}
}

\lstdefinestyle{OpenCL}{%
  language=C,
  morekeywords={size_t, uchar, ushort, uint, uchar4, char4, ushort4, short4,
    uint4, int4, float4, double4},
  keywords=[2]{__kernel, __global, __constant, __local, get_num_groups,
    get_local_size, get_group_id, get_local_id, get_global_id, get_global_size,
    barrier, mem_fence, CLK_GLOBAL_MEM_FENCE, CLK_LOCAL_MEM_FENCE,
    read_mem_fence, write_mem_fence, read_imagef, write_imagef},
  keywordstyle=[2]{\color{Cerulean}\bfseries}
}

\input{head/todonotes}                                                           
\definecolor{myblue}{RGB}{103,169,207}
\definecolor{myred}{RGB}{239,138,98}
\definecolor{mygreen}{RGB}{145,207,96}
\definecolor{myorange}{RGB}{201,147,19}
\definecolor{mycyan}{RGB}{0,177,235}
\definecolor{mygray}{RGB}{152,164,174}

\colorlet{redarea}{myred!60}
\colorlet{greenarea}{mygreen!60}
\colorlet{bluearea}{myblue!60}

\DeclareFieldFormat[inproceedings]{pagetotal}{\mkpagetotal[pagination]{#1}}
\renewbibmacro*{chapter+pages}{%
  \printfield{chapter}%
  \setunit{\bibpagespunct}%
  \printfield{pages}%
  \setunit{\bibpagespunct}%
  \printfield{pagetotal}%
  \newunit}

\DeclareFieldFormat[article]{pagetotal}{\mkpagetotal[pagination]{#1}}
\renewbibmacro*{note+pages}{%
  \printfield{note}%
  \setunit{\bibpagespunct}%
  \printfield{pages}%
  \setunit{\bibpagespunct}%
  \printfield{pagetotal}%
  \newunit}

\DeclareFieldFormat{titlecase}{\MakeTitleCase{#1}}
\newrobustcmd{\MakeTitleCase}[1]{%
  \ifthenelse{\ifcurrentfield{booktitle}\OR\ifcurrentfield{booksubtitle}%
    \OR\ifcurrentfield{maintitle}\OR\ifcurrentfield{mainsubtitle}%
    \OR\ifcurrentfield{journaltitle}\OR\ifcurrentfield{journalsubtitle}%
    \OR\ifcurrentfield{issuetitle}\OR\ifcurrentfield{issuesubtitle}%
    \OR\ifentrytype{book}\OR\ifentrytype{mvbook}\OR\ifentrytype{bookinbook}%
    \OR\ifentrytype{booklet}\OR\ifentrytype{suppbook}%
    \OR\ifentrytype{collection}\OR\ifentrytype{mvcollection}%
    \OR\ifentrytype{suppcollection}\OR\ifentrytype{manual}%
    \OR\ifentrytype{periodical}\OR\ifentrytype{suppperiodical}%
    \OR\ifentrytype{proceedings}\OR\ifentrytype{mvproceedings}%
    \OR\ifentrytype{reference}\OR\ifentrytype{mvreference}%
    \OR\ifentrytype{report}\OR\ifentrytype{thesis}}
    {#1}
    {\MakeSentenceCase{#1}}}
                                                            
\newacro{AES}{Advanced Encryption Standard}
\newacro{ALU}{Arithmetic Logic Unit}
\newacro{ASIC}{Application-Specific Integrated Circuit}
\newacro{ASIP}{Application-Specific Instruction Set Processors}
\newacro{BRAM}{Block RAM}
\newacro{CBC}{Cipher Block Chaining}
\newacro{CID}{Card Identification number} 
\newacro{CRC}{Cyclic Redundancy Check}
\newacro{CSD}{Card-Specific Data} 
\newacro{CKDF}{Concatenation Key Derivation Function} 
\newacro{DFG}{Data-Flow Graph}
\newacro{DMA}{Direct Memory Access}
\newacro{DSE}{Design Space Exploration}
\newacro{DPA}{Differential Power Analysis}
\newacro{DSP}{Digital Signal Processor}
\newacro{DUT}{Device Under Test}
\newacro{ECB}{Electronic Code Book}
\newacro{ECC}{Error-Correcting Code}
\newacro{PROM}{Programmable Read-only Memory}
\newacro{ESL}{Electronic System Level}
\newacro{FF}{Flip-Flop}
\newacro{FFT}{Fast-Fourier Transform}
\newacro{FIFO}{First Input First Output}
\newacro{FPGA}{Field-Programmable Gate Array}
\newacro{FSM}{Finite State Machine}
\newacro{FSBL}{First Stage Bootloader}
\newacro{SASB}{Self-Authenticating Secure Boot}
\newacro{SSBL}{Second Stage Bootloader}
\newacro{SSD}{Solid-State Drive}
\newacro{FPU}{Floating-Point Unit}
\newacro{GPP}{General-Purpose Processor}
\newacro{GPIO}{General-Purpose I/O}
\newacro{GPU}{Graphics Processing Unit}
\newacro{HDL}{Hardware Description Language}
\newacro{HMAC}{Keyed-Hash Message Authentication Code}
\newacro{HSM}{Hardware Security Module}
\newacro{ICAP}{Internal Configuration Access Port }
\newacro{ILA}{Integrated Logic Analyzer}
\newacro{IP}{Intellectual Property}
\newacro{IoT}{Internet of Things}
\newacro{LUT}{Lookup Table}
\newacro{MOEA}{Multiobjective Evolutionary Algorithm}
\newacro{NVM}{Non-Volatile Memory}
\newacro{PSoC}{Programmable System-on-Chip}
\newacro{PL}{Programmable Logic}
\newacro{PLL}{Phase-locked loop}
\newacro{PS}{Processing System}
\newacro{PUF}{Physical Unclonable Function}
\newacro{PRR}{Partial Reconfigurable Region}
\newacro{RCA}{Relative Card Address}
\newacro{RFS}{Root File System}
\newacro{RTL}{Register-Transfer Level}
\newacro{RLD}{Run-length Decoding}
\newacro{RSA}{Rivest, Shamir, Adleman}
\newacro{SD card}{Secure Digital memory card}
\newacro{SDIO}{Secure Digital Input/Output}
\newacro{SDF}{Synchronous Data-Flow}
\newacro{SDL}{System Description Language}
\newacro{SFU}{Special Function Unit}
\newacro{SHA}{Secure Hash Algorithm}
\newacro{SoC}{System-on-Chip}
\newacro{TLC}{Target Language Compiler}
\newacro{TMIU}{Trusted Memory-Interface Unit}
\newacro{TPM}{Trusted Platform Module}
\newacro{UML}{Unified Modeling Language}
\newacro{OS}{Operating System}
\newacro{OTP}{One Time Programmable}
\newacro{MBR}{Master Boot Record}
\newacro{MMU}{Memory Management Unit}
\newacro{DAG}{Directed Acylic Graph}

\begin{document}
\renewcommand{\baselinestretch}{1.08}

\newcommand{\CopyrightNotice}[2]{%
  \begin{picture}(0,0)(0,0)
    \put(#1){\parbox{\paperwidth-4em}{\sf \center {\small%
      The work is accepted for publication in \emph{IEEE International Symposium on Hardware Oriented Security and Trust (HOST), San Jose, USA, December 06-09, 2020}
      \textcopyright 2020 IEEE\@.
      Personal use of this material is permitted.
      Permission from IEEE must be obtained for all other uses, in any current or future media, including reprinting/republishing this material for advertising or promotional purposes, creating new collective works, for resale or redistribution to servers or lists, or reuse of any copyrighted component of this work in other works.}}}%
  \end{picture}
  \vspace{#2}
}

\title{Secure Boot from Non-Volatile Memory for Programmable SoC Architectures}

\ifx\final\undefined%
\author{\IEEEauthorblockN{\small{blinded}}
\IEEEauthorblockA{\small{blinded} \\
\small{Email: blinded}}
}
\else%
\author{\IEEEauthorblockN{\parbox{\linewidth}{\centering\small{%
                                 Franz-Josef Streit\IEEEauthorrefmark{1},
                                 Florian Fritz\IEEEauthorrefmark{1},
                                 Andreas Becher\IEEEauthorrefmark{1},
                                 Stefan Wildermann\IEEEauthorrefmark{1},
                                 Stefan Werner\IEEEauthorrefmark{2},
                                 Martin Schmidt-Korth\IEEEauthorrefmark{2},
                                 Michael Pschyklenk\IEEEauthorrefmark{2},
                                 Jürgen Teich\IEEEauthorrefmark{1}}}}%

\IEEEauthorblockA{\IEEEauthorrefmark{1}\small{Department of Computer Science, Friedrich-Alexander-Universität Erlangen-Nürnberg (FAU), Germany}}
\IEEEauthorblockA{\IEEEauthorrefmark{2}\small{Schaeffler Technologies AG \& Co. KG, Germany} \\
\small{Email: \{franz-josef.streit, florian.fritz, andreas.becher, stefan.wildermann, juergen.teich\}@fau.de} \\
\small{\{s.werner, schmmrt, pschymch\}@schaeffler.com}}
}
\fi

\maketitle

\CopyrightNotice{-40,200}{0em}
\begin{abstract}
  In modern embedded systems, the trust in comprehensive security standards all along the product life cycle has become an increasingly important access-to-market requirement.
  However, these security standards rely on mandatory immunity assumptions such as the integrity and authenticity of an initial system configuration typically loaded from \ac{NVM}. 
  This applies especially to FPGA-based \ac{PSoC} architectures, since object codes as well as configuration data easily exceed the capacity of a secure boot ROM\@.
  In this context, an attacker could try to alter the content of the \ac{NVM} device in order to manipulate the system.
  The \ac{PSoC} therefore relies on the integrity of the \ac{NVM} particularly at boot-time.
  In this paper, we propose a methodology for securely booting from an \ac{NVM} in a potentially unsecure environment by exploiting the reconfigurable logic of the FPGA\@.
  Here, the FPGA serves as a secure anchor point by performing required integrity and authenticity verifications prior to the configuration and execution of any user application loaded from the \ac{NVM} on the \ac{PSoC}.
  The proposed secure boot process is based on the following assumptions and steps: 
  1) The boot configuration is stored on a fully encrypted \ac{SD card} or alternatively Flash acting as \ac{NVM}. 
  2) At boot time, a hardware design called \ac{TMIU} is loaded to verify first the authenticity of the deployed \ac{NVM} and then after decryption the integrity of its content.
  To demonstrate the practicability of our approach, we integrated the methodology into the vendor-specific secure boot process of a Xilinx Zynq \ac{PSoC} and evaluated the design objectives performance, power and resource costs.
\end{abstract}


%
%
\section{Introduction}\label{sec:intro}
Many applications in emerging domains like the \ac{IoT}, industrial and automotive control as well as medical data processing depend on rigorous trustworthiness as well as system integrity.
However, the fact that the deployment of these embedded systems in the field requires to control/program them remotely makes them susceptible to malicious manipulation and data or \ac{IP} theft.
In order to cope with changing requirements over the lifetime of a product, connected and upgradable FPGA-based \acf{PSoC} architectures are gaining more and more interest and visibility.
Examples are platforms for the Industrial \ac{IoT}, autonomous driving, multi-camera surveillance, cloud computing, connected health and other applications~\cite{streit2018psoc,xilinxDaimler,Streit:2017,firestone2018azure,zhai2017ecg}. 
Especially here, undermining the security can have significant negative impact including monetary deficits, loss of privacy, and even physical damages.

This paper addresses the challenges of protecting the integrity and authenticity of such systems during boot, but also operational mode in a potentially unsecure environment. 
Since a processor typically boots from \acf{NVM}, e.g., a Flash memory or an external \ac{SD card}, trust can only be assumed in case that this \ac{NVM} can be a) unambiguously identified and b) checked that the content has not been altered by an attacker.
Moreover, such \ac{NVM} also often stores not only software, but also the configuration data of proprietary hardware \ac{IP} blocks to be loaded to the reconfigurable fabric of an FPGA\@. 
In addition, on many devices, the memory is used to store sensitive user data, as well as login credentials which also must be protected. 
Nevertheless, an adversary with physical access could copy or even modify the content of the \ac{NVM} in order to steal these valuable data, execute malicious code, or load undesirable hardware configurations~\cite{naumovich2003preventing}.
In this context, researchers have just recently discovered unpatchable security flaws on Xilinx Zynq UltraScale+ \acp{PSoC}~\cite{socVulnerability}.
Here, a certain secure boot mode does not authenticate the boot image metadata and partition header tables stored on \ac{NVM}, which leaves this data vulnerable to malicious modifications.
As a consequence, an attacker could modify the boot header and partition tables to load arbitrary code, thereby bypassing the entire security measures offered by the vendor's secure boot process.
Other attacks have shown that particularly configuration bitstreams could be manipulated in a way that the system exposes security sensitive information such as user data or keys~\cite{drimer2008volatile,chakraborty2013hardware}.

To address these problems and to ensure the confidentiality and integrity of the entire system configuration, a strong scheme for authentication of \ac{NVM} and its content that extends beyond existing methods is required.
In this paper, we propose therefore a hardware-centric boot process for \acp{PSoC} from \ac{NVM}. 
Its central component realized in reconfigurable logic is called \emph{\acl{TMIU}} placed as an intermediate instance between the processor and the \ac{NVM} to guarantee integrity, confidentiality, and authenticity of hardware/software programmable SoC configurations over the whole product lifetime, see~\Cref{fig:system}.
The main features of this hardware unit are: 
\begin{itemize}
  \item Entire \ac{NVM} authentication as well as sector-wise de- and encryption including metadata.
  \item On-the-fly key generation for symmetric encryption instead of external key storage.
  \item Instant integrity checking of the boot image before any user application program is loaded into volatile memory.
\end{itemize}
 
This offers several advantages over conventional software-only solutions and can build upon existing methods provided by \ac{PSoC} vendors resulting in a depth layered security. 
For instance, changing security requirements during the product lifetime can be addressed by upgrading the deployed cryptographic primitives, while software vulnerabilities and performance or power constraints are targeted through isolated execution in dedicated hardware accelerators.
In addition, without the need for external key exchange, the \ac{TMIU} allows to tie the \ac{NVM} content to a specific \ac{PSoC} device. 
In this way, these components form a permanent and immutable system to protect proprietary \ac{IP} and sensitive data.
In consequence, a configuration is only bootable from an \ac{NVM} when it has been successfully authenticated and found in an unaltered condition.
Subsequently, this raises both the trust and security level of the entire system and allows the system designer to keep track of the delivered \ac{IP}.

In the following, we define the steps and flow of this secure boot process and the concept and structure of the \ac{TMIU}. 
Subsequently, we integrate this concept into the secure boot process on a Xilinx Zynq~\emph{\ac{PSoC}}-platform containing an Artix-7-based~\emph{\ac{PL}} and a dual-core ARM Cortex-A9~\emph{\ac{PS}}.
Finally, we analyze performance as well as power and resource overhead of the proposed protection mechanisms.
The remaining of the paper is organized as follows: \Cref{sec:related} presents related work. 
Our system and threat model is briefly introduced in \Cref{sec:attack}. 
The proposed protection process and the hardware design of the \ac{TMIU} concept is described in~\Cref{sec:arch}.
Finally, experimental results of the secure boot methodology are demonstrated in~\Cref{sec:results}, followed by a conclusion and outlook on future work in~\Cref{sec:conclusion}.

%
%
\section{Related Work}\label{sec:related}
In the recent past, several mechanisms have been proposed to provide trustworthy operation of embedded systems on the one side and the prevention of \ac{IP} theft on the other side.
These mechanisms include the deployment of cryptographic hardware, secure boot, the implementation of \acp{PUF}, and techniques for \ac{IP} protection~\cite{eisenbarth2007reconfigurable,wollinger2004security,pocklassery2018self,aarestad2013help,trimberger2014fpga}.
In this context, FPGAs have been proven to offer all necessary means for protection of confidentiality, integrity, and authenticity.
Nevertheless, volatile FPGAs have an Achilles heel: they are highly vulnerable to bitstream manipulation, so-called tamper attacks, which can cause unwanted and illegal behavior or even lead to the leakage of sensitive data such as \ac{IP} and secret keys.

Previous work has also shown that reverse-engineering of \ac{LUT} content from an unencrypted FPGA bitstream is possible~\cite{ziener2006identifying}.
As a countermeasure, the FPGA vendors provide anti-tamper techniques such as symmetric bitstream encryption (AES-256) and private/public key authentication (HMAC + SHA-256, RSA-2048) already since a very long time.
By encrypting the bitstream, the design is protected against any attempt to clone or reverse engineer valuable \ac{IP}~\cite{trimberger2007trusted}. 
Whereas, bitstream encryption must be combined with authentication to protect the device against manipulation (\eg fault and Trojan injection) and to ensure that the bitstream comes from a trusted authority~\cite{drimer2007authentication}.
On Xilinx FPGAs, this can be accomplished by available hardwired on-chip crypto modules, while the necessary decryption keys can be stored on either one-time programmable E-Fuse registers or battery-backed RAM~\cite{trimberger2014fpga}.
However, both the AES and the SHA modules are not accessible from the programmable logic, and therefore, not applicable for any custom cryptographic system design.
Moreover, neither hardwired instances nor on-chip key storage can provide absolute security guarantees.
For example, in the work of Moradi et al.~\cite{moradi2012black,moradi2011vulnerability}, a side-channel attack is described that performs \ac{DPA} during bitstream encryption to extract the secret AES key, while Skorobogatov~\cite{skorobogatov2010flash} proposes optical fault injection attacks to extract decryption keys from secure embedded memory.
In~\cite{ziener2018configuration}, a fault injection attack on an FPGA is described by tampering of configuration bits inside \acp{BRAM} to extract the AES key.
Countermeasures targeting key theft can be realized by \ac{PUF} implementations to generate FPGA-internal keys. 
However, these rely either on the reverse engineering complexity of undocumented bitstreams~\cite{schellekens2008embedded} or trivialise negative effects on their reliability caused by aging and harsh environment conditions~\cite{maiti2011impact,tehranipoor2015dram}.
In comparison, our approach builds in the first instance on the basic capabilities already provided by commercial FPGAs and adds an additional but necessary layer of security by verifying the deployed \ac{NVM} and the use of on-the-fly key generation to minimize the risk of system exposure.
Moreover, the required cryptographic primitives are fully embedded into the programmable logic, and therefore, allow for customization if the security demands will change. 

Although dedicated hardware units with cryptographic engines such as \acp{TPM} and \acp{HSM} are widely used on conventional processor-based systems, cf.~\cite{wolf2011design}, and also concepts for their combination with FPGAs exists~\cite{eisenbarth2007reconfigurable}, the requirements on highly integrated \ac{PSoC} architectures differ. 
The initialization of both hardware and software after power-up is a unique feature of \acp{PSoC}.  
For this reason, the \ac{PSoC} vendors provide the option to initialize a system in a secure boot mode.
However, previous \ac{PSoC} generations used the embedded components (processors and peripherals) as slave devices to the FPGA, whereas for instance, the widely used Xilinx Zynq device family now uses the processing system as the master and the programmable logic as the slave~\cite{sanders2013secure}.
This leads to several problems, software attacks such as code injection to trigger buffer overflows or hardware Trojans to apply direct memory or bus manipulation threaten now the whole system security.
In addition, the Xilinx secure boot process is based on the creation of a chain of trust with the processor loading the entire system configuration from \ac{NVM} while relying on the confidentiality of the decryption keys stored on-board.
Although the system may take the verification and encryption of the bitstream into account, the system cannot detect or prevent any manipulation of the boot image metadata or partition tables of the \ac{NVM}, which leaves this data vulnerable to malicious modifications~\cite{socVulnerability}.

Indeed, the secure boot process on Xilinx Zynq \ac{PSoC} systems can be bypassed.
This was demonstrated in~\cite{jacob2017break} by direct memory manipulation through malicious hardware insertion.
Here, the authors exploit the vulnerability of the processor-centric architecture of Xilinx \acp{PSoC} to circumvent the secure boot process by adding an \ac{IP} block with direct memory access to the FPGA configuration.
At the moment the processor loads the bitstream from \ac{NVM} to initialize the FPGA fabric, the malicious \ac{IP} block begins to scan the main memory for the boot parameters while the processor is still booting.
As soon as these parameters are found, they are modified in a way that the processor loads an unauthorized software image from a remote server over the network instead of continuing the boot from \ac{NVM}.
In fact, such an attack can only be successful if the processor already loaded and executed code to provide the required network capabilities.
The authors provide countermeasures against this form of attack by applying a hardware wrapper for \ac{IP} blocks to prevent unauthorized memory access.
Other solutions exist in the form of hardware sandboxes as proposed in~\cite{hategekimana2017secure}.
However, our following approach is fundamentally different in specifying a hardware-centric secure boot process, where the loading of tampered configurations from \ac{NVM} to volatile memory is prevented by direct hardware authentication.

An approach similar to us addressing the secure boot on \acp{PSoC}, is \ac{SASB} proposed by Pocklassery et al.\ in~\cite{pocklassery2018self}.
Here, first an unencrypted bitstream is loaded from \ac{NVM} to implement a \ac{PUF} architecture on the FPGA\@.
In a second step, challenges are applied to the \ac{PUF} to generate a device unique key and to perform self-authentication of the loaded bitstream.
Afterwards, the key is used to decrypt the user application for the unused portion of the FPGA as well as software that runs on the processor.
The authors claim that the secure boot process is protected in a way that any modification made to the unencrypted bitstream results in key regeneration failure of the \ac{PUF}.
Potential drawbacks of this method are supposably significant resource requirements of the \ac{SASB} implementation and the fact that not only the initial bitstream but also the required \ac{FSBL} is stored unencrypted on the device, which in consequence can be tampered.
Moreover, neither the deployed device can be locked to its intended configuration nor is the configuration verified if the \ac{NVM} memory gets manipulated after the boot was successful.
Instead, we leverage existing vendor techniques by making use of the device ID provided internally by an FPGA and a unique ID of the \ac{NVM} to form an unseparable unit of device and configuration.
The integrity of swapped data after the secure boot is in our approach verified by calculating a hash for every sector exchanged between the processor and the \ac{NVM}.
Furthermore, as we will show, our solution considers not only the integrity but also the confidentiality of data during operation through hardware-based full memory encryption including the \ac{FSBL}.

\section{System and Attack Model}\label{sec:attack}
\begin{figure}[t]
  \centering
  \scalebox{1.1}{\input{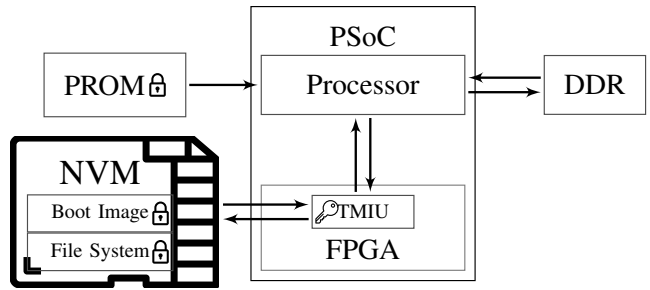}}
  \caption{Proposed security architecture involving a fully encrypted \ac{SD card} as \acf{NVM} storing the boot image as well as a file system from which a \acf{PSoC} is booted. 
  The \ac{NVM} interacts with the processor over a proposed \acf{TMIU}, which is loaded from a one-time programmable ROM (PROM) into the programmable logic of the FPGA.}
  \label{fig:system}
\end{figure}

Our system architecture is any common \ac{PSoC} as illustrated in \Cref{fig:system}, where an SRAM-based FPGA is tightly coupled with one or multiple processors integrated on a single chip.
Both the processor and the FPGA have access to an external main memory (DDR), where data is transferred after successful boot.
Yet, the data exposure of this memory by, for instance, pin probing is not considered in this work.
Instead, we consider the security vulnerabilities caused by access to \ac{NVM}.
Since SRAM-based FPGAs employ volatile memory, our approach expects the system to start the boot process from a small one-time programmable ROM (PROM) to provide an initial FPGA configuration after power-up.
The PROM on the \ac{PSoC} must be programmed ahead of shipment to contain the encrypted and signed \ac{TMIU} bitstream.
This bitstream is loaded initially into a dedicated area of the FPGA to serve as a trusted anchor between the processor and an \ac{NVM} device from which subsequently, the operating system (kernel images and device trees), partial bitstreams, and even entire file systems are booted. 

\begin{figure*}[t!]
  \centering
  \resizebox{.78\textwidth}{!}{\begin{tikzpicture} [
  auto,
  decision/.style = {diamond, draw=black, thick, fill=white, text width=5em, text badly centered, inner sep=1pt, rounded corners },
  block/.style    = {rectangle, draw=black, thick, fill=white, text width=10em, text centered, rounded corners, minimum height=2em},
  line/.style     = {draw, thick, -latex, shorten <=1pt, shorten >=1pt},
  ]
  \tikzstyle{every node}=[font=\large]
  \matrix[column sep=30mm, row sep=7mm]{%
  \node (leftHWLabel) {};                                            &                                                          & \node (rightHWLabel) {};                                           & \node[text centered] (SWLabel) {};                           \\
  \node[text centered](one){\textbf{1: Authenticate PSoC}};          & \node[text centered](two){\textbf{2: Authenticate NVM}}; & \node[text centered](three){\textbf{3: Authenticate NVM Content}}; & \node[text centered](four){\textbf{4: Grant System Access}}; \\
  \node[text centered](poweron){Power on};                           & \node[block](dongle){NVM Identification};                & \node[block](keygen){Generate AES Key};                           & \node[block](startCPU){Setup Remaining System};              \\
  \node[block](flash){Initialize TMIU (bitstream loaded from PROM)};                 &                                                          & \node[block](decryptboot){Decrypt/Hash Boot Image};                & \node[block](boot){Boot Linux Kernel (OS)};                                 \\
  \node[decision](device){$ID_{dev.}$ Match?};                       & \node[decision](cid){$ID_{NVM}$ Match?};                 & \node[decision](auth){$T_{auth.}$ \\Match?};                      &                                                              \\
  \node[block](lock0){Secure Lockdown};                              & \node[block](lock1){Secure Lockdown};                    & \node[block](lock2){Secure Lockdown};                              & \node[block](app){Launch Applications};                      \\
};

  \path[line,<->] (leftHWLabel) edge node [above] {\Large Hardware} (rightHWLabel);
  \path[line,<->] ([xshift=-5em]SWLabel.west) -- node [above] {\Large Software} ([xshift=5em]SWLabel.east);
    \path[line] (auth)       --++  (2.2,0) node [near start] {yes} |- (startCPU);
    \path[line] (poweron)     edge (flash);
    \path[line] (flash)       edge node [right] {$ID_{dev.}$} (device);
    \path[line] (device)     --++  (2.2,0) node [near start] {yes} |- (dongle);
    \path[line] (device)      edge node [right] {no} (lock0);
    \path[line] (dongle)      edge node [right] {$ID_{NVM}$} (cid);
    \path[line] (cid)        --++  (2.5,0) node [near start] {yes} |- (keygen);
    \path[line] (cid)         edge node [right] {no} (lock1);
    \path[line] (keygen)      edge node [right] {$K_{AES}$} (decryptboot);
    \path[line] (decryptboot) edge node [right] {$T_{auth.}$} (auth);
    \path[line] (auth)        edge node [right] {no} (lock2);
    \path[line] (startCPU)    edge (boot);
    \path[line] (boot)        edge (app);
\end{tikzpicture}}
  \caption{Flow diagram of the proposed hardware-centric 4-stage secure boot process from \acf{NVM}.}
  \label{fig:boot}
\end{figure*}
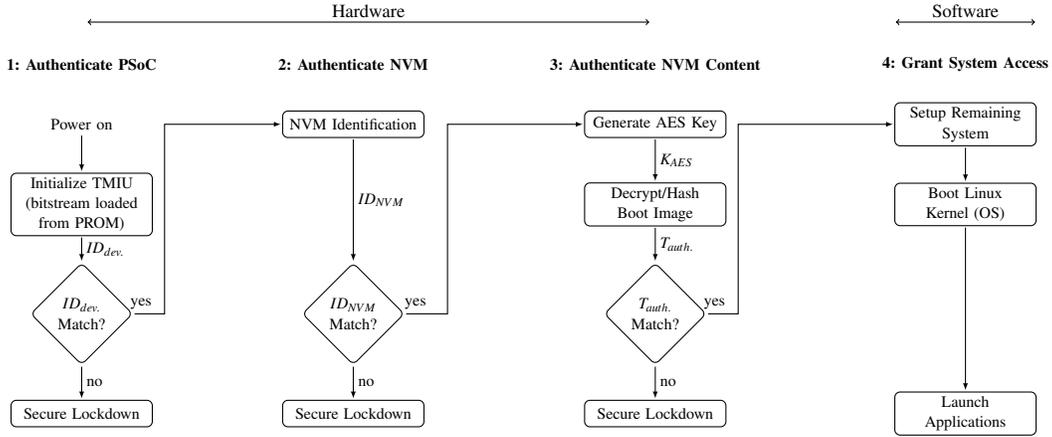

As for the attack model, we assume that the goal of an attacker is to access or modify proprietary and sensitive data stored on the \ac{NVM}.
This includes \ac{IP} in the form of object code intended to run on the processor, full or partial hardware designs, or sensitive user data. 
Furthermore, it is assumed that the adversary has full physical access to the \ac{NVM} device and can monitor and modify all communication lines to/from the \ac{NVM}.
Invasive attacks, including destructive methods to manipulate FPGA internals, system or component package damages, as well as denial of-service attacks are not taken into account. 
Again, attacks on the unprotected main memory for which countermeasures exist, \eg~\cite{gotzfried2016ramcrypt}, are not in our scope. 
Side channel attacks such as the analysis of power traces collected from the cryptographic operations performed inside the \ac{TMIU} are in the scope of future work.
We mainly address the integrity and confidentiality of the system's data stored on \ac{NVM} and, based on this, conclude a secure boot process and operation.
Moreover, we assume that the \ac{NVM} content is fully encrypted using a \ac{PSoC}-internally generated key and shipped together with the device. 
The initial configuration of the \ac{NVM} is created and transferred encrypted by the \ac{PSoC} without the used key leaving the device.
In case the \ac{NVM} should be stolen, the information on the \ac{NVM} still remains confidential due to encryption, but the \ac{PSoC} will not boot -- as desired. 
Further details of the key generation process are discussed in~\Cref{subsec:keygen}.

%
%

\section{Proposed Secure Boot Process}\label{sec:arch}
This section describes the secure boot process for \acp{PSoC} from an encrypted mass storage \ac{NVM} device.
The general approach is based on the idea of an isolated execution of security-critical operations in the reconfigurable logic of the FPGA to guarantee a trustworthy system.

\subsection{Boot Process}\label{subsec:boot}
Figure~\ref{fig:boot} illustrates the proposed 4-stage boot process providing a hardware-based chain of trust to the \ac{NVM} that includes the boot image of the \ac{PSoC}. 
This 4-stage process is explained in the following.
In stage 1), after power-on, the \acf{TMIU} according to \Cref{fig:system} is loaded from the PROM to a designated area of the reconfigurable logic region of the FPGA\@. 
The \ac{TMIU} architecture itself will be described in details in \Cref{subsec:tmiu}, see also \Cref{fig:tmiu}.

The \ac{PSoC} is authenticated by a unique, non-volatile device identifier denoted as $ID_{dev.}$. 
For $ID_{dev.}$ we make use of a 57-bit, read-only, unique board-level-identifier that Xilinx calls Device DNA~\cite{trimberger2014fpga}.
This Device DNA is burned into an E-Fuse register during the manufacturing process and can only be internally read by the FPGA design (JTAG needs to be disabled ahead of shipment). 
Similar mechanisms are offered also by other FPGA vendors such as Intel/Altera over their ALTCHIP\_ID Port~\cite{intelChipID}.
In our implementation, the \ac{TMIU} is reading out this register after power on.
If the identifier is not matching a cryptographic hash compiled into the \ac{TMIU} bitstream itself, the system will go into a secure lockdown mode.
Other essential peripherals for booting from \ac{NVM} such as clocks, \acp{GPIO}, and memory controller are also initialized at this stage.
Last in stage 1), to enable the communication between the \ac{NVM} and the processor of the \ac{PSoC}, a irreducible minimum of software is executed to establish the \ac{NVM} communication protocol. 

Once the \ac{TMIU} has been loaded, initialized, and the \ac{PSoC} device is authenticated successfully, the system goes into a memory identification mode (cf. \Cref{fig:boot} stage 2) to authenticate the connected \ac{NVM}.
In our following experiments, we use an \ac{SD card} as non-volatile mass storage device. 
This card identifies itself via its unique 128-bit \ac{CID}, which is common praxis for memory identification and exist also for Flash and other \ac{NVM} devices.
The \ac{CID}, which is factory-stamped and unchangeable, is encoded in the cards internal registers.
For this reason, as long as the card complies to the official SD standard, the \ac{CID} value denoted as $ID_{NVM}$ can be used to lock the \ac{PSoC} with device identifier $ID_{dev.}$ uniquely together with the \ac{SD card} with identifier $ID_{NVM}$.
If one does not trust the \ac{CID} alone, then for instance, the \ac{CSD} register or other card internal information can be used in combination for authentication.
The \ac{TMIU} compares the $ID_{NVM}$ number with a reference checksum also compiled into the bitstream to determine whether it is safe to activate the memory's data transfer function. 
In the situation where a different \ac{NVM} is deployed, the $ID_{NVM}$ is consequently not matching with the calculated checksum.
In this case, all I/O functions of the \ac{NVM} are suspended and the overall system will go into a secure lockdown mode. 

The third stage, also implemented fully in hardware, involves the key generation with subsequent boot image decryption and authentication.
Here, the memory identifier $ID_{NVM}$ is used together with the internally-read device identifier $ID_{dev.}$ to perform on-the-fly key generation.
As it will be discussed more detailed in~\Cref{subsec:keygen}, this guarantees a secure generation of the secret AES key denoted as $K_{AES}$ to decrypt the boot image and other data stored on the \ac{NVM}. 
After decryption, two additional authentication checks are performed.
Due to the fact, that we assume full encryption of data on the \ac{NVM}, we authenticate in a first step the encrypted \ac{MBR} of the \ac{NVM} including partition tables and other metadata.
The MBR is widely used and has established itself as the de facto standard partition table for storage media of all kinds.
In a second step, the integrity of the entire boot image is checked by means of a SHA digest calculation. 
This is done by padding the unencrypted boot image in a secure environment to a suitable message length and appending the corresponding hash token denoted as $T_{auth.}$, before storing it encrypted on the \ac{NVM}.
At boot time and after successful decryption of the overall boot image, the calculated hash is compared against the appended one.
In this way, any tampering of the encrypted boot image will be detected through a different digest value, which in turn, will lead to an immediate termination of the boot process.  
Moreover, after successful boot, the system stays upgradable at any time for secure remote updates/upgrades performed by a trusted authority.
After an authenticated remote login, a boot image could then even be replaced by a newer version on the boot partition of the \ac{NVM}.
In that case, after an automatic reboot cycle, the system would load the new configuration.
Alternatively, the old boot image could be kept as a backup if space permits and serve as a fallback in an event, where the new boot image could not be loaded as expected.
Thus, the entire system remains flexible and reduces in consequence the total costs of ownership.

We assume that the boot image can either contain one or multiple partial bitstreams to initialize the unutilized FPGA logic in combination with a bare-metal processor application or a \ac{SSBL} such as U-Boot\footnote{https://www.denx.de/wiki/U-Boot/} to deploy a Linux-based \ac{OS}. 
Persistent data storage requires a second data partition to mount the appropriate encrypted file system.
Not only during the boot process, but also during the normal operational mode of the system, the \ac{TMIU} will retain the full memory en-/decryption and hash calculation, for instance, when data is swapped to/from main memory.

Stage 4: Only if the boot partition has been successfully loaded, the \ac{TMIU} will hand over control to the processor system on the \ac{PSoC} to setup the remaining operating system and finally starts the user application. 

In summary, if any stage stops legitimization, the system will go into a secure lockdown mode, which prevents unintended system behavior and data release.
Hereafter, we describe the key generation scheme and the implementation of the \ac{TMIU} in detail.

\subsection{Key Generation Scheme}\label{subsec:keygen}
\begin{figure*}[t]
  \centering
  \scalebox{.64}{\tikzstyle{box}=[align=center,text width=7.5em,draw=black!80,fill=white,line width=.1em,rounded corners=.4em, minimum size=2em, font=\Large]
\tikzstyle{mem}=[align=center,text width=4em,draw=black!80,fill=white,line width=.1em, minimum size=2em, font=\large]

\tikzset{XOR/.style={draw,circle,append after command={
        [shorten >=\pgflinewidth, shorten <=\pgflinewidth,]
        (\tikzlastnode.north) edge (\tikzlastnode.south)
        (\tikzlastnode.east) edge (\tikzlastnode.west)
        }
    }
}

\tikzset{
  multiplexer/.style={
    draw,
    trapezium,
    shape border uses incircle, 
    shape border rotate=270,
    minimum size=18pt
  }  
}

\tikzset{
  line/.style={
  draw,
  thick,
  -latex',
  shorten <=1bp,
  shorten >=1bp
  }
}

\begin{tikzpicture}[auto] 

  \node[draw,circle, anchor=west] (VERID) {= ?};
  \node[mem,above=.5 of VERID,anchor=south] (SDID) {$ID_{NVM}$};
  \node[XOR,right=4 of VERID,scale=1.2,anchor=east] (XOR2) {};
  \node[mem,above=.7 of XOR2,anchor=south] (DNA) {$ID_{dev.}$};
  \node[box,below=1.5 of XOR2,anchor=north] (AES) {$AES$};
  \node[box,left=.5 of AES,anchor=east] (CRCIN) {$CRC_{calc./check}$};
  \node[box,right=.5 of AES,anchor=west] (SHA) {$SHA$};
  \node[draw,circle,above=1.3 of SHA, anchor=south] (VERHASH) {= ?};
  \node[box,right=.5 of SHA, anchor=west] (CRCOUT) {$CRC_{calc./check}$};
  

  
  \node[left=1.5 of CRCIN] (MUX1) {};
  \node[right=.5 of CRCOUT] (MUX2) {};

  \node[box,above=4 of MUX1,anchor=south] (CMD) {\texttt{NVM\_CMD Controller}};
  \node[left=1.5 of CMD,anchor=east] (INCMD) {};
  \node[right=15 of CMD,anchor=west] (OUTCMD) {};


  \node(KEYGEN)[dashed, draw=black!50, minimum width=8em, fit={($(SDID.west)+(0,-20pt)$) (DNA) ($(VERID.south)+(0,-5pt)$)}] {};
  \node(DATAGEN)[dashed, draw=black!50, minimum width=8em, fit={(MUX1) ($(CRCIN.north)+(0,3pt)$) ($(AES.south)+(0,-15pt)$) ($(CRCOUT.east)+(2pt,0)$)}] {};
  \node[right=.2 of VERHASH.east, text width=8em] (LabelBIT) {\large Boot Image Authentication}; 
  \node(TMIU)[draw=black!50,fit={($(INCMD.west)+(10pt,0)$) ($(CMD.north)+(0,4pt)$) ($(OUTCMD.east)+(-10pt,0)$) ($(AES.south)+(0,-32pt)$)}] {};
  
  \node[label={[rotate=90]center:\large{Non-Volatile Memory (NVM)}}, left=.1 of TMIU] (SDCARD) {};
  \node[label={[rotate=90]center:\large{Processor}}, right=.1 of TMIU] (PS) {};

  \node[above=-.15 of TMIU.south] (LabelTMIU) {\LARGE Trusted Memory-Interface Unit (TMIU)}; 
  \node[above=0 of KEYGEN.south] (LabelKeyGen) {\large \texttt{Key Generator}}; 
  
  \node[above=0 of DATAGEN.south] (LabelSDData) {\large \texttt{NVM\_DATA Controller}}; 
  \node[left=4.6 of CRCIN,anchor=east] (INDAT) {};
  \node[right=1.7 of CRCOUT,anchor=west] (OUTDAT) {};
  
  \draw[line,<->] (INCMD) edge node[above] {\large CMD} (CMD);
  \draw[line,<->] ([yshift=2ex]CMD.east) -- node[above] {\large CMD} ([yshift=2ex]OUTCMD.west);
  \draw[line,<->] (INDAT) edge node[above, xshift=-5ex] {\large DATA} (CRCIN);
  \draw[line,<->] (CRCOUT) edge node[above] {\large DATA} (OUTDAT);
  \draw[line,<->] (CRCIN) edge (AES);
  \draw[line,<->] (AES) edge (SHA);
  \draw[line,<->] (SHA) edge (CRCOUT);

  \draw[line,->] (CMD.south) -- node[left, text width=5em, xshift=1em] {\large Grant/ Block Access} ([yshift=1.4em]MUX1.north); 
  \draw[line,->] ([xshift=1em]CMD.south) |- node[right, yshift=2ex] {\large $CID$} (VERID.west);
  \draw[line,->] (KEYGEN.north) |- node[right] {\large Key Gen. Status} ([yshift=-2ex]CMD.east);
  \draw[line,->] (VERHASH.north) |- node[right] {\large Auth. Status} (CMD.east);
  \draw[line,->] (XOR2) edge node[right, yshift=-1ex] {\large $K_{AES}$} ([yshift=.7em]AES.north);
  \draw[line,->] ([yshift=.7em]SHA.north) -- (VERHASH.south);

  
  \path[line] (VERID) edge (XOR2)
  (SDID) edge (VERID)
  (DNA) edge (XOR2);

\end{tikzpicture}}
  \caption{\acf{TMIU} building blocks enabling a secure boot from external \acf{NVM}.}
  \label{fig:tmiu}
\end{figure*}
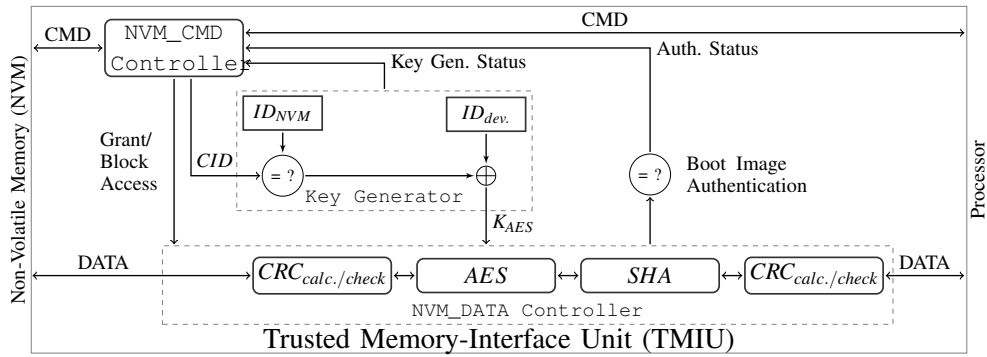

As mentioned in~\Cref{sec:related}, FPGA vendors provide on-chip key storage for bitstream decryption.
To hamper the risk of key theft, we generate the keys for decrypting the boot image and other data stored on the \ac{NVM} only after authentication, \eg only if the combination of a \ac{PSoC} with the device identifier $ID_{dev.}$ and the \ac{NVM} identifier $ID_{NVM}$ are successfully authenticated. 
This on-the-fly key generation is performed by applying the \ac{CKDF}~\cite{sp800-56a}.
As unique salt we utilize the memory identifier $ID_{NVM}$ denoted as the byte-string $OtherInfo$ and concatenate this together with the secret device identifier $ID_{dev.}$ and a counter value $c$.
In this way, the \ac{CKDF} uses the SHA implementation embedded in the \ac{TMIU} (cf. \Cref{fig:tmiu}) to calculate a pseudorandom Hash-function $H()$ for key derivation.
Here, the \ac{CKDF} is applied as a cryptographic secure key expansion function to derive the required 128-bit key for the AES from the 57-bit $ID_{dev.}$, while at the same time the difficulty of a brute force attack increases by the number of iterations specified trough $c$.

The output of this calculation is an FPGA-internal unique secret, based on the external memory identifier $ID_{NVM}$ and the private device identifier $ID_{dev.}$ that can be used as a symmetric de-/encryption key.
Equation (\ref{eq:key}) shows this derivation to generate the secret AES key $K_{AES}$.

An alternative to the vendor-provided device IDs could be to rely, for some use cases, on a \ac{PUF} implementation for device authentication as proposed in~\cite{aarestad2013help,tehranipoor2015dram}.
However, in addition to extra costs in terms of FPGA resources for a proof of concept implementation, we decided to go for the vendor ID. 

\begin{equation}
  K_{AES} = H(c  ||  ID_{dev.}  ||  OtherInfo)
  \label{eq:key}
\end{equation}
\vspace{.2cm}

The \ac{TMIU} reads the unique device identifier at the time power is applied and compares this number to a reference checksum compiled into the \ac{TMIU} bitstream.
The same applies to the \ac{NVM} identifier $ID_{dev.}$, at the memory identification stage.
So even if an attacker could read out the cards internal \ac{CID} register or probes the communication signals, no plaintext attack on the key is feasible, because the attacker has no possibility to obtain the FPGA internal $ID_{dev.}$ from a locked (\ie JTAG disabled) device.
In addition, the keys are only applied within the programmable logic of the \ac{PSoC}. 
Furthermore, a reset or power shutdown clears all generated keys and information.
It follows that the card or any other connected memory device behaves similar to a \emph{passive dongle} to tie permanently and immutable proprietary \ac{IP} to an authorized device.
Replacing either the memory device or changing its content or connecting a different \ac{PSoC} device would result in a key generation failure and prevent the system from booting.
Moreover, the \ac{NVM} itself is only read- and writable in combination with the intended \ac{PSoC} device.
In this context, also different key handling or diversification procedures are conceivable. 
For instance, multiple encryption keys could be used across multiple memory partitions to further reduce the attack surface of the system. 

\subsection{Trusted Memory Interface Unit}\label{subsec:tmiu}
To get a better understanding of how the \ac{TMIU} interacts with the \ac{NVM} device, which is in our case the \ac{SD card}, first a brief introduction of the required \ac{SDIO} protocol specification is given.
The \ac{SDIO} protocol is the standard not only for removable \acp{SD card} but also for their on-board embedded counterpart the embedded Multi Media card (eMM card) and applies the master/slave principle, where a host controller communicates over two dedicated interfaces with the memory.
A data bus is used to read or write data from the host to the memory on a bidirectional data interface (DATA).
Here, data transfers to/from the \ac{SD card} are performed in a single or multi-block read/write fashion.
A block represents the data of a specific memory sector of 512 bytes on the card and is always followed by its 16-bit polynomial \ac{CRC} value. 
This block-based communication allows our sector-wise de-/encryption and hash calculation of the memory sections.
In addition, data transfers are triggered after the controller issues specific commands over a bidirectional command interface (CMD).
These commands regulate the execution of protocol routines based on responses received from the memories internal controller. 
Similar to the data bus, every command sequence is protected by a 7-bit \ac{CRC}. 
The case that the calculated \ac{CRC} value, either for commands or data transmissions, does not match the attached one, triggers a repetition which adds an additional layer of security to the protocol.

The \ac{TMIU} is integrated as an intermediate instance to monitor and control the communication behavior between memory and host.
Figure~\ref{fig:tmiu} provides a high-level description of its building blocks.
As can be seen, the ports for the data and command lines are the only access points of the \ac{TMIU} interfacing the processor and the connected \ac{NVM}.
On the right hand side, the ports towards the processor system are shown.
The commands sent and received by the processor are observed by the component \texttt{NVM\_CMD Controller} within the \ac{TMIU}.
This controller serves not only for authentication but also to regulate the communication sequence between the processor system on the \ac{PSoC} and the \ac{NVM}.
More precisely, if any tampering on either memory or processor side occurs, the \texttt{NVM\_CMD Controller} would intervene and immediately terminate the transmission. 
As mentioned earlier, a deployed \ac{SD card} needs to authenticate itself via its \ac{CID} during the memory identification stage at boot-time.
As soon as the card sends the \ac{CID}, the \texttt{NVM\_CMD Controller} proofs and forwards this value to a \texttt{Key Generator} module.
The \texttt{Key Generator} checks whether the received memory identifier $ID_{NVM}$ and the internally-read device identifier $ID_{dev.}$ complies with the pre-initialized checksums and reports this back to the \texttt{NVM\_CMD Controller} before triggering the generation of the 128-bit AES key $K_{AES}$.
After successfully finishing the memory initialization and authentication step, the \texttt{NVM\_CMD Controller} permits the processor to transfer data and keeps track of the accessed sector numbers.

If not carefully designed, the \ac{TMIU} with its cryptographic operations would have the potential to heavily slow down not only the boot process but also general system performance.
As high data throughput from memory to \ac{PSoC} and vice versa is a must, special care has been taken in the implementation of the dedicated \texttt{NVM\_DATA Controller}.
In particular, all internal components are interconnected by a streaming interface.
This interface is designed for high-speed data throughput and supports burst transmissions of unlimited size. 
Therefore, no address mechanism or explicit synchronization is needed, which makes it ideal for pipelined data streaming.
However, incoming data from the memory or the processor must first be proven non-faulty, \eg against any transmission errors (\eg bit flips).
This is done by \ac{CRC} calculation before data is forwarded to the AES de-/encryption module.
If the \ac{CRC} check is not successful, the data is sent unencrypted to the \ac{PSoC} to trigger a \ac{CRC} error resulting in a new transmission.
After de-/encryption, the \ac{CRC} value is calculated again and subsequently attached to the data block before forwarding it to the PSoC/\ac{NVM}.

Our AES core supports sector-wise both de- and encryption and is switching between these two modes depending on the communication direction.
The key which was used initially for encrypting the memory content in the secure environment ahead of shipment is now also applied when writing/reading data to/from \ac{NVM} at boot-time and during the operational mode of the system.
In this way, we can guarantee that at any point in time, only encrypted data is stored on \ac{NVM}.
Moreover, the \ac{NVM} content is only deployable on the intended \ac{PSoC} device.

The \texttt{NVM\_DATA Controller} utilizes the 256-bit SHA algorithm to compare the calculated boot image hash token against the sent one $T_{auth.}$.
If the calculated hash is not matching the appended one, the \texttt{NVM\_DATA Controller} modifies the last byte to provoke a \ac{CRC} error on the processor side to invalidate the data transmission, while the \texttt{NVM\_CMD Controller} disables any further attempts to transfer data. 
In a similar fashion, sector authentication is performed after successful booting, when reading or writing data to the \ac{NVM}.
In this way, every sector is stored with a hash and fully encrypted on the \ac{NVM} making any manipulation immediately detectable.
Later, when a reset or power-off is applied, all internal registers are cleared and the \ac{TMIU} will again start the proposed multi-stage boot process.

Finally, the \ac{TMIU} is supposed to be clearly separated from the user design, while the remaining non-configured regions are free and can be used for user applications. 
This gets possible through the partial reconfiguration and isolated place and route capabilities of modern FPGAs~\cite{xilinxFences}.
In particular, it allows us to dynamically perform partial reconfigurations of certain independent regions to update the desired system functionality when the system is applied in the field.
Therefore, the \ac{TMIU} with all its security sensitive primitives is logically and spatially isolated and runs totally independent from other applications.

%
%
\section{Experimental Results}\label{sec:results}
\begin{figure}[t]
  \centering
  \scalebox{1}{\input{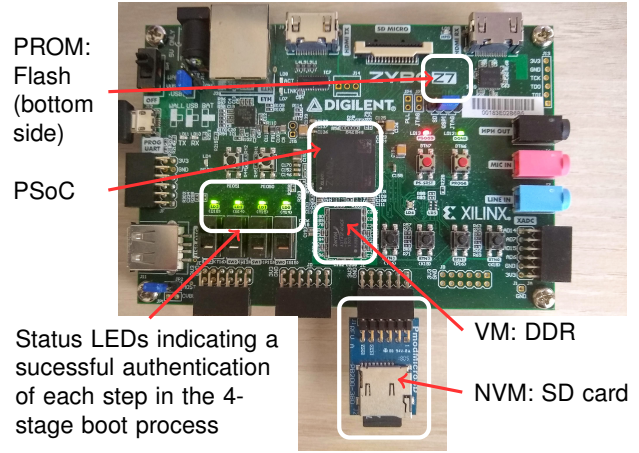}}
  \caption{Secure boot setup on a Xilinx Zynq evaluation board.}
  \label{fig:board}
\end{figure}

In this section, our proposed hardware-centric boot process and \ac{TMIU} design is evaluated on a Xilinx Zynq xc7z010clg400-1 placed on the Digilent Zybo evaluation board, see~\Cref{fig:board}.
For a proof of concept, an \ac{SD card} has been chosen as non-volatile mass data storage device for the following reasons:
SD as well as eMM cards are cost effective due to their high memory density and good cost-per-bit ratio, making them suitable for storing relatively large amounts of data such as user data.
Moreover, they can be easily used in a number of small, lightweight and low-cost systems due to their low-power consumption and standard sizes, which is a common requirement for many embedded and \ac{IoT} devices. 
Nevertheless, the approach is not restricted to \acp{SD card}.
Also different forms of \ac{NVM} with similar properties such as \acp{SSD} or Flash chips are possible.

In order to successfully boot from a Zynq device, the experimental setup illustrated in \cref{fig:board} was required:
As a PROM which would allow to store the bitstream of our \ac{TMIU} that needs to be loaded initially is not available on the Zynq, we used an on-board (bottom side) available \SI{17}{\mega\byte} programmable Flash.
This Flash is connected over a SPI-Flash controller with the processing system to emulate our PROM and assumptions introduced in~\Cref{sec:attack}.
The storage requirement for the encrypted \ac{TMIU} implementation is \SI{1.9}{\mega\byte} in total, with the bitstream accounting for the major part of the storage size.
Beside the Flash memory, a \SI{16}{\giga\byte} \ac{SD card} was connected via an external MicroSD card slot to the \ac{PSoC}.
The card was formatted with one \SI{100}{\mega\byte} FAT32 bootable partition containing the required boot image, device tree and the Linux kernel image.
The remaining card space was utilized as ext4 partition to provide the Linux root file system.
Furthermore, in our evaluation the boot process involves the configuration of the processor and the hardware.
Therefore, the boot image includes a partial FPGA configuration in combination with the \acf{SSBL} (U-Boot in this case) to load the kernel image and setup a Linux OS\@.
Next, the processing system on the Zynq is routed via its \ac{SDIO} interface to the \ac{TMIU} ports on the FPGA\@. 
The clock speed of the \ac{TMIU} adapts during the memory initialization phase according to the connected card standard. 
In this work, we were using a high speed card clocked by \SI{50}{\MHz}.
To visualize the in \Cref{fig:boot} proposed 4-step boot process, four status LEDS were used indicating every successful authentication.

Furthermore, the Xilinx Vivado Design suite 2018.1 was utilized to synthesize the \ac{TMIU} on the aforementioned Zynq \ac{PSoC}.
Here, the design objectives performance, power, and resource costs were collected at a \SI{20}{\nano\second} clock cycle time.
The amount of required FPGA resources for the \ac{TMIU} implementation in terms of \acp{FF}, \acfp{LUT}, and 36K~\acp{BRAM} -- no \ac{DSP} 48-slices were needed -- is highlighted in \Cref{tab:res_tmiu}.
The numbers indicate that, with 39\% of \acp{LUT} and 17\% of \acp{FF} on the second smallest \ac{PSoC} from the Zynq family, a \ac{TMIU} implementation is even feasible on entry-level \ac{PSoC} devices.

Concerning power consumption, an estimation including both static and dynamic power of the overall design including the \ac{TMIU} netlist was obtained using the Vivado Power analysis tool.
The switching activity was derived from constraints and simulation files.
As a result, considering the additional FPGA resources introduced by the \ac{TMIU}, the overall power consumption of the design rises from 1.53 to 1.62 Watt, which is an overhead of only 5\%.
In this overall power, the processor consumes about 92\% of the dynamic on-chip power, while clock and register activity of the \ac{TMIU} contributes to 8\% of the dynamic power.
The static device power amounts to 7\% of the overall power consumption.


\begin{table}[b]
  \caption{Resource requirements of the proposed \ac{TMIU} implementation and its cryptographic building blocks.}
  \label{tab:res_tmiu}
  \centering
  \resizebox{\columnwidth}{!}{%
  \begin{tabular}{llrrr}
    \toprule
      Block                                                    & Num.          &   LUTs        &    FFs        & BRAMs\\
    \midrule                                                                                                                 
    \multirow{2}{*}{\texttt{Key Generator}}                    & abs.          &  70           &  249          &   0  \\
                                                               & (\%)          &  0.4          &  0.7          &   0  \\
    \cmidrule{2-5}                                                                                                           
    \multirow{2}{*}{\texttt{NVM\_CMD Controller}}              & abs.          &  83           &  331          &   0  \\
                                                               & (\%)          &  0.5          &  0.9          &   0  \\
    \cmidrule{2-5}                                                                                                           
    \multirow{2}{*}{\texttt{NVM\_DATA Controller}}             & abs.          & 6630          & 5354          & 0.5  \\
                                                               & (\%)          & 37.7          & 15.2          & 0.8  \\
    \cmidrule{1-5}                                                                             
    \multirow{2}{*}{\textbf{Total}}                            & \textbf{abs.} & \textbf{6783} & \textbf{5934} & \textbf{0.5}\\ 
                                                               & \textbf{(\%)} & \textbf{38.5} & \textbf{16.9} & \textbf{0.8}\\ 
    \bottomrule
  \end{tabular}
  }
\end{table}

\begin{figure}[t]
  \hspace{-3.5em}
  \centering
  \scalebox{.53}{\newlength{\xdim}

\definecolor{myred}{HTML}{D7191C}
\definecolor{myorange}{HTML}{FDAE61}
\definecolor{mygreen}{HTML}{228B22}
\definecolor{myblue}{HTML}{2B83BA}

\pgfplotsset{compat=newest} 

\begin{tikzpicture}
\begin{axis}[
    xbar stacked,
    tick label style={font=\Large},
    label style={font=\Large},
    axis lines*= left,
    axis x line*=bottom,
    xtick={0,100,200,300,400,500,600,700},
    x label style={at={(axis description cs:.5,-.4)},anchor=north},
    xlabel={Time in ms},
    xmin=0,
    xmax=700,
    width=.9\textwidth,
    bar width=7mm,
    axis y line*=none,
    enlarge y limits=0.48,
    symbolic y coords={Secure Boot},
    yticklabel style={text width=2cm,align=right},
    ytick distance=2,
    tick align = outside,
    y=10mm,
    legend style={%
    legend columns=2,
        at={(xticklabel cs:0.5)},
        anchor=north,
        draw=none
    },
    legend style={at={(axis description cs:.5,1.8)},anchor=north, font=\Large},
    area legend,
]

\addplot[myblue,fill=myblue!20] coordinates {(98,Secure Boot) (0,Secure Boot)};
\addplot[mygreen,fill=mygreen!20] coordinates {(0,Secure Boot) (624,Secure Boot)};

\legend{PROM, SD-Card}

\coordinate (throughputPROM) at (axis cs:49,Secure Boot);
\coordinate (throughputSD) at (axis cs:350,Secure Boot);


\coordinate (startPROM) at (axis cs:0,Secure Boot);                                          
\coordinate (endPROM) at (axis cs:98,Secure Boot);

\coordinate (startSD) at (axis cs:98,Secure Boot);                                          
\coordinate (endSD) at (axis cs:624,Secure Boot);

\end{axis}  

\node at (throughputPROM) {\large\SI{19.4}{\mega\byte\per\second}};
\node at (throughputSD) {\Large\SI{24.7}{\mega\byte\per\second}};

\draw [decorate,decoration={brace,amplitude=10pt,raise=12pt}] (startPROM.north west) -- (endPROM.north east) node [black,midway, xshift=-2mm,yshift=13mm] {\Large{TMIU Bitstream = \SI{1.9}{\mega\byte}}};
\draw [decorate,decoration={brace,amplitude=10pt,raise=12pt}] (startSD.north west) -- (endSD.north east) node [black,midway, xshift=8mm, yshift=12.6mm] {\Large{Boot Image + Linux Kernel + Device Tree = \SI{13}{\mega\byte}}};
\end{tikzpicture}}
  \caption{Performance of the \acf{TMIU}-based secure boot.
  In the first phase, the bitstream of the \ac{TMIU} is loaded and initialized from PROM\@. 
  Subsequently, the boot process continues from an authenticated \ac{NVM} (in this case an \ac{SD card}). 
  As can be seen, the data rate achieved is close to the \SI{25}{\mega\byte\per\second} line rate of the \ac{SD card}.}
  \label{fig:latency}
\end{figure}
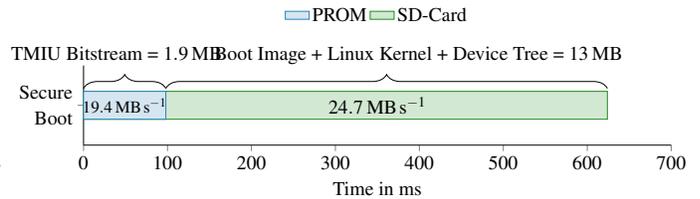

In the following, we evaluate the boot time and achievable data transmission rate between \ac{NVM} and \ac{PSoC} using the proposed \ac{TMIU} concept and implementation.
Figure~\ref{fig:latency} shows the time needed for loading the \ac{TMIU} configuration and the \acf{FSBL} from a PROM including the time for \ac{TMIU} initialization and mutual device authentication (steps 1-3 in \Cref{fig:boot}) which amounts to \SI{98}{\milli\second} corresponding to a throughput of \SI{19.4}{\mega\byte\per\second}.
Thereafter, the \ac{TMIU} continues booting the Linux OS from the \ac{NVM}.
In our experiment, data with a total size of \SI{13}{\mega\byte} for the required boot image, device tree and the Linux kernel image is loaded.
The time for this subsequent boot from an \ac{SD card} specified to provide a maximal line rate of \SI{25}{\mega\byte\per\second} was measured as \SI{526}{\milli\second} corresponding to a data rate of \SI{24.7}{\mega\byte\per\second}.
Therefore, the \ac{TMIU} design does not reduce the achievable throughput of the \ac{NVM} device and perfectly scales with the amount of data loaded at boot time.
Concerning latency, the decryption and authentication of data in hardware takes 52 clock cycles to process a sector of size 512 Bytes.
Therefore, the boot time is limited only by the bandwidth of the chosen \ac{NVM} medium and its interface. 
Furthermore, resource, power, and timing overheads caused by the proposed \ac{TMIU} approach are tolerable for the sake of security.
An evaluation of alternative \ac{NVM} devices is subject of future work.

\section{Conclusion and Future Work}\label{sec:conclusion}
In this paper, we presented a novel approach for checking authenticity of \acf{NVM} and the integrity of stored boot images for programmable \acf{SoC} devices. 
Here, apart from boot ROMs, \ac{NVM} devices such as SD and eMM cards or Flash memories are typically used due to their high capacity.  
In order to prevent any fraudulent exchange of the memory device or modification of its content, a fully hardware-centric solution is proposed in which a so-called \acf{TMIU} is loaded first into the available reconfigurable region of the FPGA, which then initializes the communication interface of the \ac{PSoC} with the \ac{NVM} device including authentication, integrity checking, and de-/encryption of data. 
A 4-stage boot process and its hardware implementation have been evaluated in terms of resource utilization, power, and performance.  
As a result, the lightweight, low-power \ac{TMIU} implementation can be used already in quite small, even \ac{IoT} devices. 
Moreover, the proposed protocol has been shown to not limiting the speed of the boot process from \ac{NVM}.
Additionally, external key storage for decryption is avoided through on-the-fly key generation by making use of unique, factory-stamped IDs given by the device and the \ac{NVM}.
Due to the fact that our \ac{TMIU} design and proposed protocols do not use any vendor-specific hardware or software primitives, our approach can target any \ac{PSoC} platforms integrating processor and FPGA resources on a chip.

In future work, we want to investigate more deeply the applicability of the proposed approach in combination with dynamic update/upgrade services and their requirement for secure off-chip storage including policies for fail-safe fallback modes if these updates fail.
Furthermore, it is planned to investigate the vulnerability of the approach to side-channel attacks. 
Last but not least, we intend to analyze the approach also for other types of \ac{NVM} devices. 

%
%
\ifx\final\undefined%
\section*{Acknowledgment}
blinded
\vspace{2cm}
\else%
\section*{Acknowledgment}
The work has been supported by the Schaeffler Hub for Advanced Research at Friedrich-Alexander University Erlangen-Nürnberg (SHARE at FAU).
\fi
%
%
\printbibliography
\end{document}